\documentstyle[12pt]{article}
 
\newcommand{\be}{\small\begin{equation}}
\newcommand{\ee}{\end{equation}\normalsize\vspace*{-0.1ex}}
\newcommand{\bea}{\small\begin{eqnarray}}

\newcommand{\eea}{\end{eqnarray}\normalsize\vspace*{-0.1ex}}
\newcommand{\bdm}{\small\begin{displaymath}}
\newcommand{\edm}{\end{displaymath}\normalsize\vspace*{-0.1ex}}
\newcommand{\beas}{\small\begin{eqnarray*}}

\newcommand{\eeas}{\end{eqnarray*}\normalsize\vspace*{-0.1ex}}
 
\begin{document}
 
 
\thispagestyle{empty}
\renewcommand{\thefootnote}{\fnsymbol{footnote}}
 
\setcounter{page}{0}

\begin{flushright}
UM - TH - 97 - 11\\
hep-ph/9705318
\end{flushright}

\begin{center}
\vspace*{2cm}
{\Large\bf Renormalons and $1/Q^2$ Corrections}\\
\vspace{1.8cm}
{\sc R. Akhoury and V.I. Zakharov}\footnote{Talk presented at the Conference "Beyond the standard model",
Balholm, Norway, May 1997}

Randall Laboratory of Physics,

University of Michigan  Ann Arbor, Michigan 48109, U.S.A.

\vspace*{2.0cm}
{\bf Abstract} 

\end{center}

We argue that the
appearance of the Landau pole in the running coupling of QCD
introduces $1/Q^2$ power corrections in current correlation functions.
 These terms are not accounted for by the standard operator product expansion
and is the price to be paid for the lack of a unique definition
of the running coupling at the $1/Q^2$ level. We review also possible 
phenomenological implications of the $1/Q^2$ terms in an alternative language
of the ultraviolet renormalon.

\newpage
\renewcommand{\thefootnote}{\arabic{footnote}}
\setcounter{footnote}{0}
 
 
{\bf 1.} Renormalons by construction are 
a part of the
dynamics of the standard model since they are 
simply a set of perturbative graphs existing within, say, 
QED or QCD \cite{thooft} (for a review see, e.g., Ref. \cite{az}).
Nevertheless, there are links of renormalons to the physics beyond the
standard model. For example, one may consider renormalons within 
theories going beyond the standard model, in particular, supersymmetric gauge
theories \cite{az}. In this part of the talk we shall choose another route
and concentrate on hints to the non-standard dynamics revealed by renormalons.
More specifically, we will consider $1/Q^2$ terms which are absent from
the standard operator product expansion \cite{svz} but are indicated by
the ultraviolet (UV) renormalons \cite{vz,va,yz,alt,peris}. 
The material we are presenting 
is to a great extent  
of a review nature. There are also some new points, in particular, 
the relation of the $1/Q^2$ terms
to the Landau ghost has not been emphasized so far.
  
{\bf 2.} To explain, why the $1/Q^2$ terms could signal a kind of
non-standard dynamics let us first outline schematically the standard
picture. For our purposes,
it is most coveniently formulated in terms of the QCD sum rules.
The basic feature of the sum rules is the emphasis on
the power-like corrections. In a simplified form, one derives
the sum rules of the type:
\be
\int exp(-s/M^2)R(s)ds~\approx~
(parton~model)\left(1+{\alpha_s(M^2)\over \pi}+c_G{<0|\alpha_s(G_{\mu\nu}^a)^2|0>\over M^4}+...\right)\label{sr}
\ee
 where $R(s)$ is the total cross section
of $e^+e^-$ annihilation into hadrons
in the standard units, $M^2$ is a large mass parameter and $<0|\alpha_s(G_{\mu\nu}^a)^2|0>$
is the so called gluon condensate, $c_G$ is a coefficient calculable 
perturbatively. 

Moreover, in analyzing the sum rules one assumes that it is the
power-like corrections of order $M^{-4}$ which signal the breaking of
asymptotic freedom at moderate $M^2$ of order $\le (GeV)^2$.
Phenomenologically, this breaking is due to the appearance of resonances.
Note the absence of $1/M^2$ corrections from Eq. (\ref{sr}).
This is a direct consequence of the OPE since the first gauge invariant operator, that is $\alpha_s(G_{\mu\nu}^a)^2$ has dimension $d=4$.

It is important to emphasize that 
the matrix element $<0|\alpha_s(G_{\mu\nu}^a)^2|0>$
is saturated by infrared contributions. In particular,
the $1/M^4$ corrections can be traced by means of infrared
renormalons \cite{mueller}. 
Within the renormalon approach the gluon condensate manifests itself as
an $n!$ growth of
the coefficients $a_n$ of perturbative expansions in the running coupling
$\alpha_s(Q^2)$:
\be
(a_n)_{IR}~\sim~\int_0^{\sim Q^2}k^2dk^2 (b_0)^n ln(Q^2/k^2)^n~\sim
b_0^n2^{-n}n!
.\ee
Here $n$ is the order of perturbative expansion considered to be large,
$b_0$ is the first coefficient
of the $\beta$-function which is positive in QCD,
 $k^2$ is virtual mometum flowing through gluon line and
$Q$ is an external momentum such that 
$Q^2\gg\Lambda^2_{QCD}$.
The large numerical value of $a_n$ is due to the contribution
of characteristic momenta of 
order
\be
k^2_{char}~\sim~e^{-n/2}Q^2
.\ee

Independent of whether one is using
the general OPE or IR renormalons, the resulting picture can be 
described in a very simple way: the presence of 
low-lying resonances on the
phenomenological side is signaled by power corrections of infrared narure
derivable within fundamental QCD.
 
{\bf 3.} This picture, which seems perfectly selfconsistent,
is challenged by UV renormalons.
Ultraviolet renormalons are known \cite{thooft}
to dominate perturbative expansions at large orders
of perturbation theory.
The expansion coefficient at large $n$ is proportional to:
\be
(a_n)_{UV}~\sim~
\int_{\sim ~Q^2}^{\infty} (b_0)^n {dk^2\over k^4}(ln Q^2/k^2)^n~\sim~
(-1)^nb_0^nn!\label{uv}
.\ee
Note that we will use a one-term $\beta$-function for simplicity.
In fact, evaluation of the UV renormalon can be 
pursued much further \cite{va,beneke} but the improvements are
not crucial for our purposes.

Thus, there are two basic features of the UV renormalons which are important
for our discussion:

(i) the UV renormalon is Borel summable because of the sign oscillations,
$(-1)^n$,

(ii) UV renormalons are related to very large virtual momenta:
\be
k^2_{char}~\sim~e^nQ^2.
\ee
Both features seem to meet our intuition: 
because of asymptotic freedom
the large momenta in QCD should represent no major problems and,
as a reflection of this, the contribution of the UV renormalons 
can be summed up.

However, a paradox arises \cite{vz} once it is realized
that 
the UV renormalons are Borel summable to a $1/Q^2$ piece:
\be
\sum_{n_{crit}}^{\infty}b_0^n(-1)^nn!(\alpha_s(Q^2))^n~\rightarrow~
c_{UV}{\Lambda_{QCD}^2\over Q^2}\label{uvv}\ee
where $n_{crit}$ is the critical value of the order of the perturbative
expansion starting from which the perturbative contributions start
to rise as function of $n$ because the factor $n!$ prevails over
the factor $(\alpha_s(Q^2))^n$ and
$c_{UV}$ is a constant, while the arrow means that 
the standard Borel summation is applied to the asymptotical perturbative
expansion.
 
What is surprising about the Eq. (\ref{uvv}) is that it defies the
standard picture outlined above. Indeed, phenomenologically
the $\Lambda_{QCD}^2/Q^2$ piece would be associated with the
contributions 
of low-lying resonances while the UV renormalon is associated
with large virtual momenta $k^2\gg Q^2\gg\Lambda_{QCD}^2$.
Thus, it appears that physics in the
infrared should match physics in ultraviolet
as far as power like corrections are involved. 

Another important point about the UV renormalons is that
the Borel summation is not the only way to deal with
the divergence (\ref{uv}) of the perturbative expansions
and the results of alternative procedures is not obviously the same.
Namely, one can utilize either a conformal mapping \cite{mueller1}
or expansion in the coupling normalized at a high scale $\mu^2$,
$\mu^2\gg Q^2$ \cite{bz} to avoid the divergence due to the UV renormalon. 
In particular, if one uses the expansion in $\alpha_s(\mu^2)$
then the uncertainty of the perturbative expansion due to
its asymptotical nature caused by the UV renormalons
is of order \cite{bz}:
\be
\Delta_{UV} (Q^2)~\sim ~c_{UV}{\Lambda_{QCD}^2\over Q^2} {Q^4\over \mu^4}
\ee
and can be made arbitrarily small by choosing $\mu^2\gg Q^2$.
Although this trick might appear to solve 
the problem of the UV renormalon it rather brings new problems, in fact.
Indeed, if one detects 
the presence of a $1/Q^2$ correction in one formulation of the perturbative
expansion and looses track of it while using another formulation 
then this might imply either the inconsistency of the whole approach or
the existence of further consistency conditions that have yet to be established.

{\bf 4.} The paradoxes are resolved, to our mind, by the simple observation
that perturbatively the coupling is not well defined at the $1/Q^2$ level
because of the Landau pole.
Indeed, if we use the running coupling in the standard form:
\be
\alpha_s(Q^2)~=~{1\over b_0lnQ^2/\Lambda_{QCD}^2}\label{rc}
\ee
then we introduce a pole at $Q^2=\Lambda_{QCD}^2$
which is a fake singularity in the sence that all the perturbative
graphs have only cuts which start from $s=0$. The emergence of the
Landau pole is an artifact of the summation procedure which leads to
(\ref{rc}). This summation procedure is well known 
to be justified in the leading log 
approximation and does not introduce any inconsistency at
the logarithmic level. However, as far as the power like correction 
$1/Q^2$ is
concerned it is in fact not fixed by perturbation theory itself. 
Note that our simplifying use of the one-term $\beta$-function is not
crucial at this point, at least in so far as the use of a
more realistic $\beta$-function
does not remove the Landau pole.

Thus, when one demonstrates the absence of the $1/Q^2$ corrections
by applying the expansion in $\alpha_s(\mu^2),\mu^2\rightarrow \infty$
(see discussion above) one heavily uses in fact
the analyticity properties of the perturbative graphs.
On the other hand, the use of the running coupling $\alpha_s(Q^2)$
to demonstrate the $1/Q^2$ uncertainty due to the UV renormalon
relies on a resummation procedure which does not observe these
analyticity properties at the level of the $1/Q^2$ terms.
As long as the problem of the Landau pole is not solved within 
perturbation theory there is no way to decide which
derivation is to be prefered.

One may introduced a running coupling with the Landau ghost removed:
\be
\tilde{\alpha}_s(Q^2)~=~{1\over b_0ln Q^2/\Lambda^2_{QCD}}+
{\Lambda_{QCD}^2\over b_0(\Lambda_{QCD}^2-Q^2)}\label{red}
\ee
which at large $Q^2$ clearly differs from the "standard" coupling (\ref{rc})
by a $1/Q^2$ correction. This kind of redefinition of the coupling goes back 
to the fifties \cite{redmond} and was reviewed very recently
in the context of QCD in Ref. \cite{arbuzov} where further references
can also be found. 
One can of course introduce further modifications which in turn remove 
the $1/Q^2$ term:
\be
\tilde{\alpha}^{'}_s(Q^2)~=~
{1\over b_0lnQ^2/\Lambda_{QCD}^2}+{\Lambda^2_{QCD}\over b_0(\Lambda_{QCD}^2-Q^2)}+{\Lambda^2_{QCD}\over b_0 Q^2}\label{new}\ee
According to ref. \cite{arbuzov}
the advantage of this form 
is that the pole is shifted now from unphysical $Q^2$
to $Q^2=0$ which is the beginning of the physical cut.
Moreover, the coupling (\ref{new}) differs from the standard
definition (\ref{rc}) only by $1/Q^4$ terms.

Thus, the $1/Q^2$ terms are viewed now as  
 arising from the uncertainties of the perturbative definition
of $\alpha_s$. Although the OPE says that there is no $1/Q^2$ correction
this should be understood rather as a statement about the difference of
the full answer for the polarization operator $\Pi (Q^2)$
(or its imaginary part proportional to $R(s)$) and its perturbative
expansion (in finite orders). 
Since the perturbative expansion itself is in fact not
defined at the $1/Q^2$ level, the prediction on the absence of the
$1/Q^2$ terms is not well formulated yet.
An extra hypothesis is to be made concerning the precise definition
of $\alpha_s$ which avoids $1/Q^2$ corrections in the 
polarization operator. 
\footnote{One of the present authors (V.Z.)
discussed the Landau pole as an origin of the $1/Q^2$ terms
in the coupling constant with M. Beneke and V. Braun 
in 1994. A {\it critique}
of this point of view can be found in Ref. \cite{beneke2}
where it is claimed that the Landau pole does not introduce
any uncertainty of order $1/Q^2$ in theoretical predictions
for physical quantities. It is our understanding that this statement
is based in fact on a tacit assumption that $1/Q^2$ terms
do not enter the OPE provided that the standard running coupling 
or its modification a la (\ref{new})is used. Because of the inconsistency
of the perturbation theory revealed by the Landau pole
this assumption cannot be proven, to our mind.}
Also, the interplay between low and high momenta revealed first
by the UV renormalons (see above) becomes less surprising.
Indeed, it is non-perturbative effects which presumably settle
the problem of the Landau pole in the infrared. 
The corresponding ultraviolet "tail" 
in the coupling behaves as $1/Q^2$. 

It is worth emphasizing that various definitions of the coupling are 
not simply related by adding or removing $1/Q^2$ in a universal way from
all terms of different order in $\alpha_s$.
Consider as an example a term of second order in the running coupling
$\sim \alpha^2_s(Q^2)$. Imagine furthermore that we would like 
to remove an unphysical singularity due to the Landau pole from
the dispersive representation of $\alpha_s(Q^2)$,
i.e. to work out an expression similar to Eq. (\ref{red}).
To this end we should remove a single pole from: 
\be
{1\over ln^2Q^2/\Lambda^2_{QCD}}~\rightarrow~
{1\over ln^2Q^2/\Lambda^2_{QCD}}-{\Lambda^2_{QCD}\over Q^2-\Lambda^2_{QCD}}
.\ee 
Thus, we readily see that as far as $1/Q^2$ terms are concerned
the uncertainty in $\alpha_s^2$ 
is no less than in $\alpha_s$ so that the whole perturbative 
series collapses for the power correction. A similar phenomenon occurs in fact
in case of the $1/Q$ correction due to IR renormalons in event shapes
(for a review, see, e.g., Ref. \cite{az}).

From the phenomenological point of view this collapse of the perturbative
expansion in the power-like corrections implies
that the value of the $1/Q^2$ correction is to be considered
as an independent fit parameter. In particular,
relation between $1/Q^2$ contributions to various observables can
be established only at the price of new model dependant assumptions.

{\bf 6.} The lack of guidance as to which model is to be selected
has hindered the progress in the phenomenology of $1/Q^2$ corrections
and we conclude this note with a mini review of the attempts to
develop the phenomenology of $1/Q^2$ corrections made so far.

In fact these attempts were formulated mostly in terms of the UV renormalons.
On the other hand, one may notice that the very existence of the $1/Q^2$
can be guessed simply on the basis of existence of the Landau pole.
We do not think, however, that this change of the language is indeed very significant. The central problem is working out a reasonable model and any
approach to the $1/Q^2$ terms 
would be the equally good in so far as the model turns out to be 
adequate in describing the data.

The most frequently discussed channel is the $e^+e^-$ annihilation into
hadrons, ( see in particular Refs. \cite{yz,alt,narison} ) since the 
data are most abundant here. 
Estimates of $1/Q^2$ terms in the tau-decays
which arise from using various 
types of dispersion relations can be found in Ref. \cite{alt}.
On the other hand.
one can invert the problem and get constraints on possible $1/Q^2$
terms from the data \cite{narison}. 
The bounds turn to be quite stringent.

Relatively large $1/Q^2$ would be welcome on phenomenological grounds 
in the pseudoscalar (pion) channel \cite{yz}.
Moreover, the description 
of the $1/Q^2$ corrections in terms of the UV renormalon can match in this 
case the description of low-energy physics in terms of the Nambu-Jona-Lasinio
model. While this hypothesis is far from being firmly established
let us mention developments on the theoretical side
which do favour this possibility.
First, it turns out that the UV renormalon is dominated by contribution
of the same four-quark operators which are postulated within
the NJL model \cite{va}. This is not a trivial statement since
these operators emerge first only on the level of two renormalon chains, not
a single renormalon chain. This dominance of the four-quark operators in
the UV renormalon is true for various further observables 
as well \cite{beneke}.

Although it is attractive to assume that a $1/Q^2$ correction is actually
the leading power-like correction in the pseudoscalar channel,
at first sight this picture cannot be reconciled with
the bounds on the $1/Q^2$ in the vector channel (see discussion above).
It seems indeed remarkable therefore
that the direct evaluation of the two renormalon
chains in the vector and pseudoscalar channels indeed indicates
a substantial numerical disparity of these contributions \cite{peris}.
More specifically it turns out that
the contribution of the UV renormalon in these channels is related
as  \cite{peris}:
\be
(UV~renormalon)_{PS}~=~18 ~(UV ~renormalon)_V
\ee
where the large numerical factor is the result of explicit calculations.

{\bf 7.} To summarize, the very existence of the Landau pole
induces, generally speaking, $1/Q^2$ corrections at large $Q^2$.
Alternatively, these corrections can be ascribed to the UV
renormalons. Although the phenomenology of such corrections is still
in its infancy there are some reasons to expect that the pion is
in fact dual to the $1/Q^2$ corrections of the fundamental QCD. 
\footnote{A possible connection to 
the Adler anomaly is worthy of investigation.}
If this interpretation turns to be true
it would provide with a new insight into the dynamics of hadrons and
interplay of short and large distances in QCD which might be 
useful for the understanding of the physics beyond the standard model as well.

We would like to acknowledge useful discussions with V. Braun
and A.I. Vainshtein. Very recently, similar remarks
on the role of the Landau pole in generating $1/Q^2$
terms were made by G. Marchesini and G. Grunberg
and we are thankful to G. Marchesini for private communications
on these matter. Finally, after this note was written there
appeared a paper by G. Grunberg \cite{grunberg}
devoted to the role of the
Landau singularity in generating $1/Q^2$ corrections.



\end{document}